\documentstyle[12pt,aasms4,graphics]{article}

\begin{document}

%\submitted{Submitted to ApJLett, 1998 Oct 21; revised 1998 Dec 14; Accepted 1998 Dec 21}

\title{SGR 1806-20 Is a Set of Independent Relaxation Systems.}
\author{David M. Palmer}
\affil{Universities Space Research Association, Code 661, Goddard/NASA, Greenbelt, MD
20771, palmer@lheamail.gsfc.nasa.gov}
\authoremail{palmer@lheamail.gsfc.nasa.gov}

\begin{abstract}
The Soft Gamma Repeater 1806-20 produced
patterns of bursts during its 1983 outburst that indicate multiple independent
energy accumulation sites, each driven by a continuous power source, with
sudden, incomplete releases of the accumulated energy. The strengths of the
power sources and their  durations of activity vary over 
several orders of magnitude.
\end{abstract}


\section{Introduction}

Soft Gamma Repeaters (SGRs) are very highly magnetized
($B\sim\!10^{15}\,\rm{G}$), slowly rotating ($P\sim\!8\,\rm{s}$), young
($\sim\!10^{4}\,\rm{years}$) neutron stars that produce multiple bursts of
soft gamma-rays, often  at super-Eddington luminosities
($10^{37-41.5}\,\rm{erg}$ in a few tenths of a  second). Two of these objects
(SGR 0526-66 and SGR 1900+14) have also produced hard, extremely intense
superbursts ($10^{44.5}\,\rm{erg}$ in a few tenths of a  second).
%, followed by a minutes-long softer tail which is modulated with 
%the neutron star rotation period).
In 
the Thompson \& Duncan\markcite{ThomDun} (1995) model, the smaller bursts are
produced by  `crustquakes' in the neutron star, while the larger bursts are 
produced by global reconfiguration of the magnetic field.

Four of these objects have been discovered so far, including SGR
1806-20\markcite{Atteia}\markcite{Laros87} (Atteia et al. 1987; Laros et al. 1987),
the subject of this Letter. Observations of SGR 1806-20 with the
\textit{XTE} PCA (1996 Nov) and \textit{ASCA} (1993 Oct) find that its 
quiescent emission shows a 7.47 s periodicity with a spin-down rate  of
$\dot{P} = 8 \times 10^{-11} \rm{s\,s^{-1}}$, implying
%$\dot{P} = 3 \times 10^{-11} \rm{s\,s^{-1}}$
%(PCA data alone) or $8 \times 10^{-11} \rm{s\,s^{-1}}$ (PCA and
%\textit{ASCA})\markcite{chryssa} (Kouveliotou et al. 1998). The higher
%spindown rate, measured over a longer  baseline, implies
a magnetic field of $8 \times
10^{14} \rm{G}$ and a  characteristic spin-down age $P / 2 \dot{P}$ of
$\sim$1500 years\markcite{chryssa} (Kouveliotou et al. 1998).
This source is  associated with the SNR G10.0-0.3, which has
an inferred age of $\sim$5000 years\markcite{kulkarni93} (Kulkarni \& Frail,
1993). Corbel et al. (1997)\markcite{corbel} measure the distance to this SNR
as $14.5\pm1.4$ kpc.

The intervals between successive bursts are distributed lognormally
\markcite{hurley} (Hurley et al. 1994).
Cheng et al. (1995) \markcite{cheng} found that this distribution, the
correlation between successive waiting intervals, and the distribution of
intensities (a $dN/dS \propto S^{-1.66}$ power law with a high-intensity
cutoff) are similar to the behavior of earthquakes. Previous 
analyses have found no clear relationship between the timing of the bursts and 
their intensities\markcite{ulmer}\markcite{laros87} (Laros et al. 1987; Ulmer et al.
1993).

This Letter demonstrates that SGR 1806-20 contains multiple systems that
continuously accumulate energy and discontinuously release it as bursts.  This
is consistent with the crustquake model (with multiple seismic zones) but not
with, e.g., impact event, continuous accretion, or disk instability models.

\section{Observations and Analyses}

The data analyzed in this Letter are from the 134 bursts catalogued in
Ulmer et al. (1993)\markcite{ulmer} from the University of California,
Berkeley/Los Alamos National Laboratory instrument on the
International Cometary Explorer (\textit{ICE}) during the SGR's 1979-1984 term
of activity. This activity peaked in  1983 Oct-Nov with more than 100 detected
bursts, including 20 on Nov 16 (\textit{D}ay \textit{O}f \textit{Y}ear 320).
Figure 1a is the history of bursts during this period, showing the time $t_i$
and burst size $S_i$ as measured by counts in the 26-40 keV channel of
\textit{ICE}'s scintillating detector.

Figure 1b shows the cumulative fluence, the running sum of the burst sizes, as a
function of time.  If we assume that the burst catalog provides a good and
complete measure of the energy emitted as bursts by this source, then we may
use this to understand the energetics of the bursting mechanism.  Section  2.1,
below, examines and validates this assumption.

Figure 1b shows that the rate of energy release varies
dramatically over this time period. However, intervals are apparent when the
average power, averaged over many bursts, is approximately constant, giving a constant
slope. The intervals marked $A$, $B$, $C$ and $D$ are selected
for further study in this Letter.

A \textit{relaxation system} is a system which continuously accumulates an input
quantity (e.g., energy) in a reservoir, and discontinuously releases it. For a
system that starts with a quantity $E_0$ in its reservoir, accumulates at a
rate
$R(t)$, and releases with events of size $S_i$ instantaneously at times $t_i$, the
contents of the reservoir as a function of time is given by
$$E(t) = E_0 + \int_0^tR(t')\,dt' - \sum_i^{t_i < t}S_i$$

The simplest behavior from such a system occurs with a constant
accumulation rate $R(t) = r$, a fixed `trip point' which triggers a release when
$E = E_{trip}$, and a constant release size $S_i = s$,
giving a periodic relaxation oscillator with $P = t_{i+1}-t_i = s/r$. If the
accumulation rate, trip point, or release strength are not constant, then more
complicated behavior results.  Stick-slip (including earthquake) and avalanche
systems are other examples of relaxation systems. 
If the reservoir has a maximum capacity $E_{max}$ such that
\hbox{$0 \le E \le E_{max}$}, and constant rate
$R(t) = r$, the sum of releases approximates a linear function of time:
\hbox{$\sum_i^{t_i < t}S_i = \left[(t-t_0)r\right]_{-E_{max}}^{+0}$}.
The linear sections of Figure 1b can be tested to see if they are consistent
with such relaxation systems.

The interval $B$, detailed in Figure 2a, demonstrates this behavior.  The
cumulative fluence is bounded above by a linear function
of time, corresponding to a rate of 392 counts per day.  The maximum deviation
below the line is comparable to the size of the largest bursts.  Assuming that
energy flows into a reservoir at this constant rate and is released only as the
catalogued bursts, we can model the energy in the reservoir by subtracting the
emitted fluence from the integrated input energy, as shown in Figure 2b.
%(The relationship between detector counts and source
%energy is discussed in Section 2.1.)
The bursts have an apparent tendency to
keep the reservoir at low levels---the cumulative fluence tends to stay near
the linear function.  A statistical analysis (see Section 2.2) shows that, if
these same bursts were arranged randomly, the cumulative fluence would tend to
deviate much more from the best linear rate than is observed here.  This is
very strong evidence that these bursts come from a relaxation system.

Interval $C$ may be a continuation of this relaxation system.  Its average rate
is consistent with that of $B$ and, if it is assumed that a few of the many
bursts in the intervening time are from this system, the two intervals can be
combined in a plot qualitatively similar to Figure 2.  However, the interval
between $B$ and $C$ is the most active period ever seen for this source,
including a single hour which has approximately as much fluence as the entire
5-day interval $B$.  That this violent activity does not affect the parameters
of the relaxation system suggests that it comes from a physically independent
site, perhaps a different location on the neutron star.

The 9-burst interval $A$, detailed in Figure 3, is also consistent with a
relaxation system if you omit the single burst that occurs at 1983 DOY
297.940.  The remaining bursts are consistent with a constant-$R$ relaxation
system in which most of the bursts are total releases of the reservoir energy. 
The rate for $A$ is a factor of 20 below that of
$B$ and $C$, suggesting that it is a different system.  The high statistical quality of
the relaxation system fit to those eight bursts clearly identifies the remaining burst
as an interloper (Section 2.2).

The interval $D$ (11 bursts), appears to have two different energy
accumulation rates differing by $\sim$40\% (57 counts/day for 5 bursts $D_{i}$,
then 96 counts/day for 6 bursts $D_{ii}$). These rates are comparable to each
other, and far from those of
$A$ and $B$, so $D$ may represent a single system that speeds up slightly
between $D_{i}$ and $D_{ii}$, rather than two different systems.

%Intervals $C$ and $D_{i}$ cannot be combined without at least two
%rate changes: a large decrease near the last burst of $C$ followed by an
%increase near the first burst of $D_{i}$.
%This is further evidence for multiple relaxation systems in SGR 1806-20.

\subsection{Cumulative Fluence as an Integrating Bolometer}

The sum of the catalogued burst intensities is a good measure of the
integrated burst energy emitted by the object to the extent that
a) the catalog contains all bursts above a certain threshold and is uncontaminated,
b) the bursts below that threshold contain only a small fraction of the total energy,
c) the detected counts are proportional to the energy
released in the direction of the detector, and
d) the fraction of energy released in the direction of the detector
is constant and, specifically, independent of the neutron star spin phase.
Violation of any of these conditions would cause `noise' in our analysis, which
could distort our understanding of the source's burst energy output.

\textit{a) catalog completeness}---\textit{ICE} was an interplanetary spacecraft, 
and so its observations were not continually interrupted by 
occultations, as Earth satellite observations often are, nor hampered 
by rapidly varying background from orbiting within Earth's 
magnetosphere. \textit{ICE} thus provides a long, continuous, and stable set 
of measurements resulting in a uniform catalog with good completeness down
to the  instrument's sensitivity limit of  $\sim$16 counts.  The false-trigger rate
in the catalog is estimated to be $<1\,\rm{year}^{-1}$\markcite{laros87} (Laros et
al. 1987).

\textit{b) subthreshold bursts}---The observed burst intensity distribution power law,
$S^{-1.66}$
\markcite{cheng} (Cheng et al. 1995) has an index $\gamma > -2$, which
places most of the energy in the largest bursts up to the high-E cutoff.
Recent \textit{XTE} PCA observations find that this power-law distribution
extends to bursts far below the \textit{ICE} threshold (Dieters, priv.
comm.).  An extrapolation to zero of the integrated energy as a function of
burst size indicates that
$\sim28\%$ of the burst energy is sub-threshold. The noise due to sub-threshold
bursts would be muted if the energy comes out in the form of many small bursts,
randomly distributed in time.  Indeed, 80\% of the sub-threshold energy is extrapolated
to be in bursts at least a factor of 2 below our threshold but, as this
Letter shows, bursts are not randomly distributed in time, so these
small bursts still have the potential to cause problems.

\textit{c) intensity-energy relationship}---The catalogued burst size (the
number of counts detected in the 26-40 keV channel of a scintillator) is
proportional to the total energy fluence incident on the detector if the bursts
always have similar spectra. Fenimore, Laros \& Ulmer\markcite{fen94} (1994) find
that the spectral shape of bursts from this source is largely independent of the
burst fluence with a small scatter in the burst hardness.
%There is statistically-detectable scatter in the burst hardness
%ratios which could be either intrinsic to the  source or due to
%systematic effects in the detector---either of which would degrade the
%relationship between detected counts and energy fluence.  However, these
%variations are relatively small.
The spectral fits indicate that each count
represents the emission of
$\sim4 \times 10^{38} \rm{erg}$ of X/$\gamma$-rays.

\textit{d) isotropic emission and rotational modulation}---SGRs are rotating
neutron stars, and anisotropic emission would make the relationship
between total emitted energy and the detected counts dependent on the neutron
star's rotational phase.
Fourier analysis of the times of bursts in interval $B$ (which, coming from a
single system, might be expected to show the strongest phase coherence) showed
no significant modulation for periods between 7.40 and 7.48 seconds---the
reasonable range of extrapolations to 1983 of the
Kouveliotou et al.\markcite{chryssa} (1998) measured period and spindown rate.
Weighting the times either directly or inversely with $S_i$ showed that both
strong and weak bursts were independent of spin phase.

These conditions merely ensure that the measured running sum of burst sizes is a good
approximation to the total emitted burst X/$\gamma$-ray energy.  The energy which flows
into the reservoir may escape, undetected, through other channels of non-burst or
non-X/$\gamma$ energy release.  However, as the data show, any energy leaks are
not severe enough to conceal SGR 1806-20's relaxation system behavior.

\subsection{Statistical Analysis}

Any set of events can be trivially described as a relaxation system, assuming a
sufficiently large reservoir and an arbitrary set of release times and sizes.
However, relaxation systems with specific properties can be distinguished from
random systems by use of a test statistic designed to detect those properties. If
the observed value of this test statistic is outside of the range expected for a
random process, then that is proof that the system is non-random, and evidence of
a physically-significant relaxation system.

For a relaxation system with a constant accumulation rate, a reservoir small compared to
the total emitted energy, and a tendency to release a large fraction of the available
energy, a promising statistic is the Sum Of Residuals (SOR). This is the sum of the
energies left in the reservoir immediately after each burst $SOR =
\sum_i E(t_i^+)$. Since the contents of the reservoir and the accumulation rate
are not directly observable, the SOR is minimized with respect to a constant
rate
$r$ and an empty-reservoir time $t_0$ with the constraint that each
residual
\hbox{$E(t_i^+) = (t-t_0)\,r - \sum_j^{j \le i}S_j$} must be $\ge0$ for all $i$.

The SOR value can be calculated for the observed data and then, by a bootstrap
method, compared to the distribution of SOR values calculated for randomized
versions of the data.
For this analysis, the randomized data is produced by `shuffling'
(selection without replacement) the burst intensities while keeping the burst
times the same.  This procedure trivially preserves all previously-known
characteristics of the data (intensity distribution, interval distribution,
and interval-interval correlation) to ensure that the relaxation
system behavior is not an artifact of these characteristics.

The entire 134-burst catalog was searched for the sub-interval of $N=33$
bursts with the lowest SOR.  This located the interval $B$.  Then, for each
of $10^6$ trials, the burst intensities were shuffled as described above, and
the randomized catalog was again searched for the $N$-burst interval with the
lowest SOR.  (In most cases, this interval was essentially the same as
interval $B$.)  In only 21 of $10^6$ trials was the SOR lower than that of
the observed data. (This result is moderately insensitive to the value of
$N$, giving similar values for
$N=30$ and $N=35$, and increasing by an order of magnitude at $N=25$.)
This demonstrates at the 99.998\% confidence level that the
intensities are inconsistent with chance, and are correlated with burst
times in a manner that implies a relaxation system during this time period.

With the existence of relaxation systems demonstrated, they can be sought in
other intervals, with the randomization restricted to specific sets of
consecutive bursts.
For all 9 bursts in $A$ (DOY 294.867-312.797)
the SOR statistic does
not distinguish the measurements from the randomized trials. However, when
the SOR minimization is permitted to discard any one burst (from each of the
observed and randomized trials), the observed data is non-random at the
$6.6 \times 10^{-5}$ level. This is evidence, not just that 8 of the bursts
are from a relaxation system, but that the remaining burst belongs to a
different system. The false-trigger rate in this catalog is estimated to be
$<1\,\rm{year}^{-1}$\markcite{laros87} (Laros et al. 1987), or a $<5\%$
chance of occurring during this time interval. This implies that the extra
burst was from SGR 1806-20, but was not from the system responsible for the
other bursts during $A$.
%Exhaustive bootstrap, 24 out of 9! lower 

Interval $D$, as fit by a relaxation system with a single rate change, is
non-random at the 97\% level.  This result, although marginal, suggests that
systems on SGR 1806-20 can change their accumulation rate.

\section{Discussion}

As shown in this Letter, some of the bursts from SGR 1806-20 come from
relaxation systems. Additional examples can be found in the burst history, and
parsimony suggests that all SGR bursts (except, perhaps,
superbursts) are from such systems.  This would not always be easily demonstrable,
even if observations meet all of the requirements of Section 2.1.  Single systems
that produce only a few bursts, rapidly vary their accumulation rates, or have
reservoirs large compared to the typical burst size could be indistinguishable
from random. Multiple simultaneously active systems could be difficult to
disentangle.

% 1 count is equiv to 4e38 erg, so 1 count/day = 4.6e33 erg/s

The accumulation rates for the intervals discussed in this paper are as low as
19 counts/day (equivalent to $\sim 10^{34.9}\,\rm{erg\,s^{-1}}$) for the 18-day
interval
$A$, and as high as 392 counts per day (equivalent to $\sim 10^{36.3}
\,\rm{erg\,s^{-1}}$) for interval
$B$.  Accumulation rates up to 80,000 counts/day (equivalent to
$\sim10^{38.6}\,\rm{erg\,s^{-1}}$) are seen during the peak hour of activity on
1983 Nov 16, probably in one or a few relaxation systems.  There are also
quiescent intervals when no bursts are seen for years---between
1985 Aug and 1993 Sep, the available instruments (which provide incomplete coverage)
recorded no bursts that are attributed to SGR 1806-20.

If all SGR bursts are from relaxation systems, the available data show that there can
be multiple systems active simultaneously.  The conditions which activate such
systems are global to the SGR---several systems are apparent in the months
covered by this Letter, but there are years when no systems are
active.  However, each system is independent in that it has its own
accumulation rate and reservoir.

Relaxation systems are incompatible with SGR models in which each burst is produced
by the accretion of a distinct object, such as neutron stars invading another
star's Oort cloud.  Continuous accretion with episodic burning, as in a nova, is
a type of relaxation system, but it would be difficult for accretion to
independently feed multiple sites, each with its own rate and beginning and ending
times.

This analysis extends the earthquake analogy of Thompson \&
Duncan\markcite{ThomDun} (1995) beyond the similarity of the
$S^{-1.66}$ intensity distribution and the interval relationships found by
\markcite{cheng} Cheng et al. (1995). Seismic regions are relaxation
systems, driven by quasi-steady accumulation of tectonic stress due to
continental drift, with sudden, incomplete releases of energy as
earthquakes. Tsuboi\markcite{tsuboi} (1965), using an analysis similar to
this
Letter's, found that the energy released by earthquakes in and near
Japan during 1885-1963 is consistent with a constant input rate into a
finite reservoir.

Further studies of this aspect of SGRs can be made using continuously-operating gamma-ray
instruments on interplanetary spacecraft, such as those on \textit{Ulysses},
\textit{Near Earth Asteroid Rendezvous} and
\textit{Wind}.  Although \textit{XTE} is continually interrupted by Earth
occultations, its high sensitivity might allow it to determine if the smallest,
most frequent 
bursts also demonstrate the behavior of relaxation systems over short time
intervals.  The comparison of SGRs with earthquakes may improve our understanding
of both types of events.

%\newpage

\begin{figure}
\resizebox{7in}{3.5in}{\includegraphics{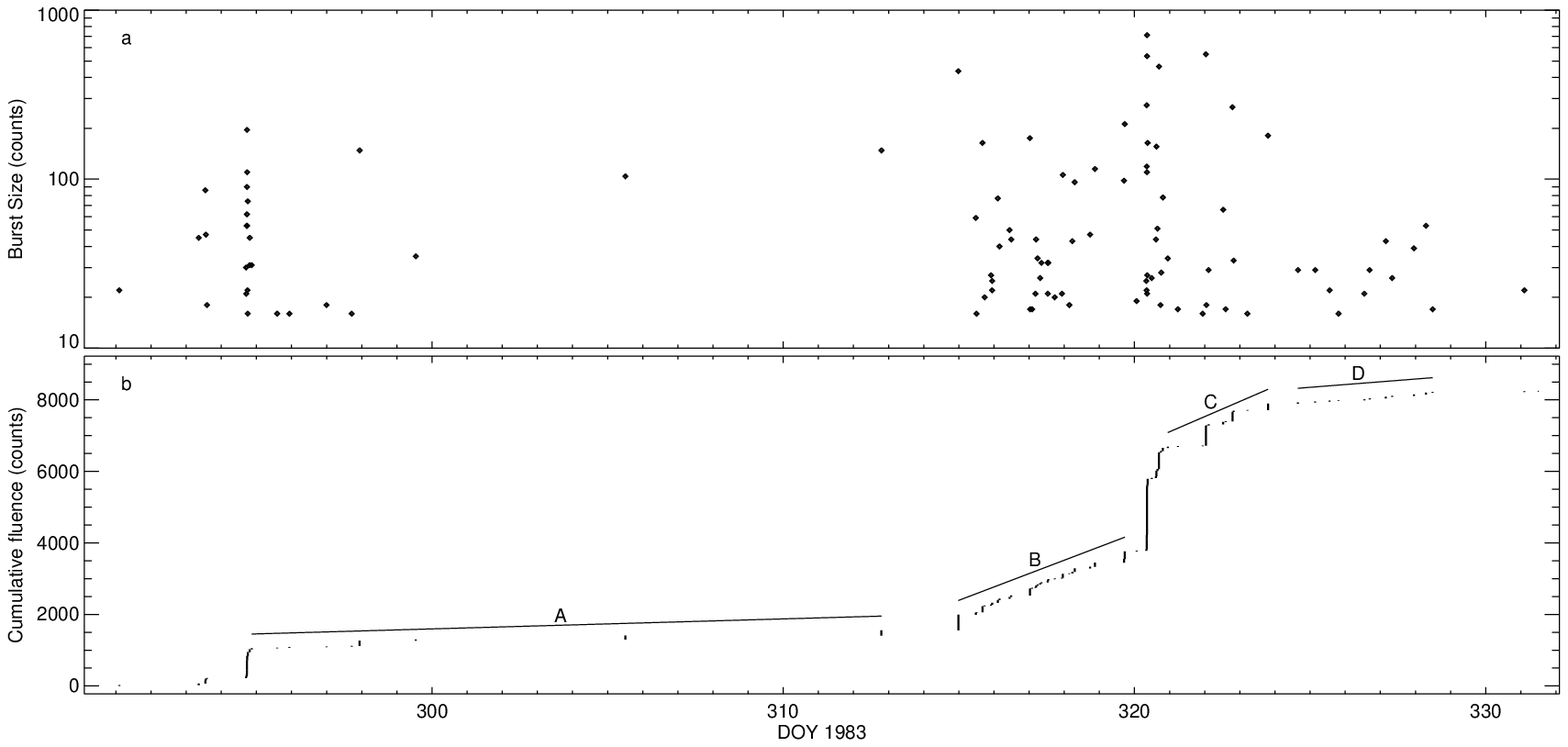}}
\figurenum{1}
\caption{\textit{a)} Catalogued SGR 1806-20 bursts from 1983 Oct-Nov (103
bursts).  Burst size is the fluence in counts from \textit{ICE}.  \textit{b)}
The cumulative burst fluence (running sum of burst sizes from Figure 1a). 
Intervals $B$ and
$A$ are detailed in Figures 2 and 3, respectively.}
\end{figure}
\begin{figure}
\resizebox{3.5in}{3in}{\includegraphics{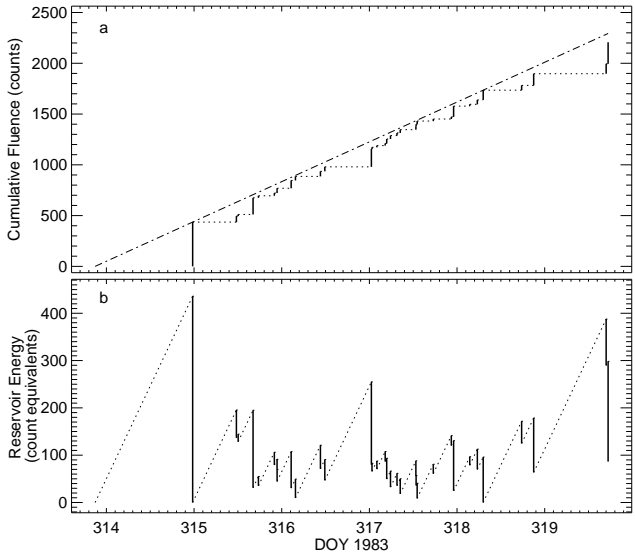}}
\figurenum{2}
\caption{\textit{a)} Cumulative burst fluence for interval $B$ (33 bursts).  Each
vertical line is an burst.  The dot-dashed line represents a steady rate of 392
counts per day. 
\textit{b)} The corresponding reservoir energy.  This is modeled assuming a
constant power input into a reservoir, which releases the energy as bursts. One count
equivalent corresponds to the emission of $\sim4\times10^{38}$ ergs of
X/$\gamma$-rays.}
\end{figure}
\begin{figure}
\resizebox{3.5in}{3in}{\includegraphics{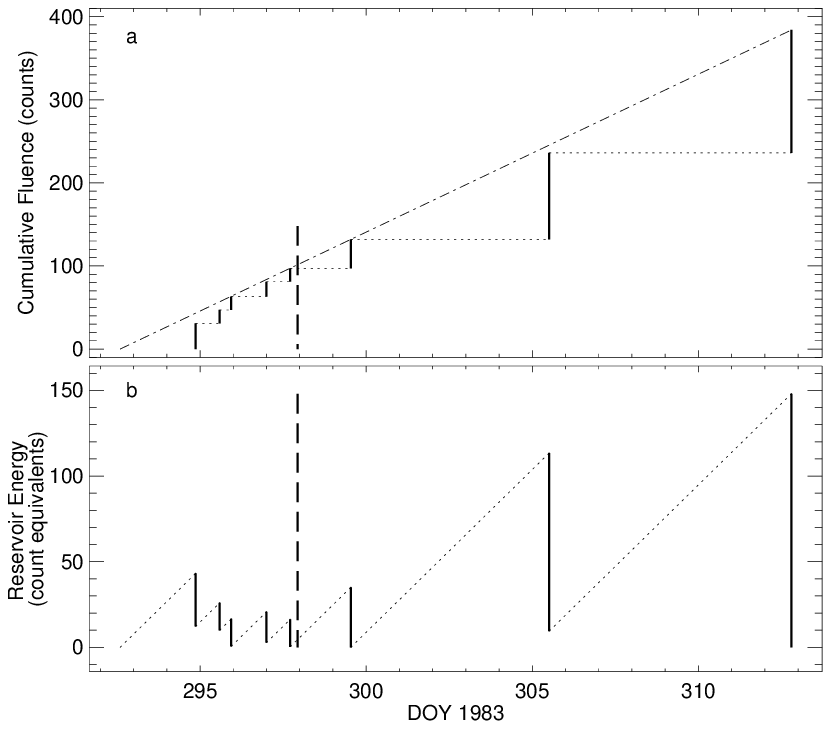}}
\figurenum{3}
\caption{\textit{a)} Cumulative burst fluence for interval $A$, exclusive of the burst
at DOY 297.940 (dashed vertical line).  The dot-dashed line represents a rate of 19
counts/day.  \textit{b)}  Modeled reservoir energy, also excluding the extraneous
burst.}
\end{figure}


\begin{references}
\reference{Atteia} Atteia, J.-L., et al., 1987, \apj, 320, L105
\reference{cheng95} Cheng, B., Epstein, R.I., Guyer, R.A., \& Young, A.C. 
1995, Nature 382, 518
\reference{corbel} Corbel, S., Wallyn, P., Dame, T.M., Duruchoux, P., Mahoney,
W.A., Vilhu, O., \& Grindlay, J.E. 1997, \apj, 478, 624
\reference{fen94} Fenimore, E.E., Laros, J. G., \& Ulmer, A. 1994,
%The X-ray spectrum of the soft gamma repeater 1806-20,
\apj, 432, 742
\reference{chryssa} Kouveliotou, C., et al. 1998, Nature, 393, 235
\reference{hurley} Hurley, K.J., McBreen, B., Rabbette, M., and Steel, S.,
1994, \aap, 288, L49
\reference{kulkarni93} Kulkarni, S.R. \& Frail, D.A., 1993, Nature, 365, 33
\reference{laros87} Laros, J.G., et al. 1987, \apj, 320, L111
\reference{ThomDun} Thompson, C., \& Duncan, R.C., 1995, \mnras, 275, 255
\reference{tsuboi} Tsuboi, C., 1965, Proc. Jap. Acad., 41, 392
\reference{ulmer} Ulmer, A., Fenimore, E.E., Epstein, R.I., Ho, C., Klebesadel,
R.W., Laros, J.G., \& Delgado, F. 1993, \apj, 418, 395
\end{references}
\end{document}